\documentclass[conference]{IEEEtran}

\usepackage{amsmath,amssymb}
\usepackage{booktabs}
\usepackage{graphicx}
\usepackage[caption=false,font=footnotesize]{subfig}
\usepackage{cite}
\usepackage{url}
\usepackage{xcolor}
\usepackage{hyperref}
\usepackage{makecell}
\usepackage{comment}

\graphicspath{{figures/}}

\newtheorem{remark}{Remark}

\title{Pilot-Based Sensing With OFDM and AFDM: A Resource-Matched Performance Comparison}

\author{
\IEEEauthorblockN{Siyi Tang\IEEEauthorrefmark{1}, 
Yin Xu\IEEEauthorrefmark{1}, 
Aimin Tang\IEEEauthorrefmark{1},
Dazhi He\IEEEauthorrefmark{1}, and
Wenjun Zhang\IEEEauthorrefmark{1}}
\IEEEauthorblockA{\IEEEauthorrefmark{1} Shanghai Jiao Tong University, Shanghai, China\\
Email: \{tangsiyi, xuyin, tangaiming, hedazhi, zhangwenjun\}@sjtu.edu.cn}
}

\begin{document}
\maketitle

\begin{abstract}
As a waveform candidate for next-generation wireless systems, affine frequency division multiplexing (AFDM) has often been compared with orthogonal frequency division multiplexing (OFDM) under different configurations, while a systematic comparison of their pilot-based sensing performance remains lacking. This paper establishes a comparison with the same bandwidth, frame duration, pilot energy, and reserved-resource overhead. OFDM-random, OFDM-fixed-comb, and OFDM-shifted-comb pilots are compared with AFDM pilots surrounded by guard resources using aperiodic ambiguity functions without cyclic prefix/chirp-periodic prefix (CP/CPP) and cyclic ambiguity functions with CP/CPP, followed by target detection experiments. The results show that OFDM-fixed-comb exhibits low range sidelobes but periodic range ambiguity peaks, while AFDM and OFDM-shifted-comb effectively suppress zero-Doppler range ambiguities. Within the common unambiguous region, all pilot configurations for both waveforms exhibit similar detection thresholds in strong-target detection, while structured OFDM and AFDM pilot placements outperform OFDM-random in weak-target detection. At high mobility, the AFDM receiver using the discrete affine Fourier transform (DAFT) can resolve integer Doppler beyond the conventional slow-time maximum unambiguous velocity, outperforming the tested conventional OFDM and AFDM receivers at the increased search complexity. These results separate the effects of waveform, pilot placement, and receiver processing, providing a basis for AFDM and OFDM pilot-based sensing design.
\end{abstract}


\section{Introduction}
Sixth-generation networks are expected to support reliable connectivity and environmental awareness, such as connected vehicles. Integrated sensing and communication (ISAC) addresses these requirements by reusing spectrum, hardware, and waveforms for both data delivery and target/environmental sensing \cite{liu2022isac,yin2025adaptive}. By utilizing the existing communication infrastructure, wireless sensing can be achieved by reusing communication signals in 6G systems. Basically, sensing may exploit pilots, decoded data, or hybrid signals. Pilot-based sensing is attractive as the deterministic pilots used for synchronization and channel estimation can provide known sensing samples without prior data decoding, maintaining compatibility with existing frame structures. Many studies have been carried out for pilot-based sensing, including experimental validations with 6G ISAC systems for multi-target tracking and behavior recognition \cite{tang20266g,wang2026enabling}.  
%With sparse pilot resources, the waveform structure and the two-dimensional pilot placement affect range--Doppler resolution, ambiguity, and detection performance.

Orthogonal frequency division multiplexing (OFDM) provides a natural baseline for ISAC systems. Its mature transceiver architecture, compatibility with current broadband standards, and fast Fourier transform (FFT) implementation enable efficient range-Doppler processing and have motivated extensive OFDM-ISAC research \cite{sturm2011fusion,mirabella2023ofdm}. 
With limited pilot resources, pilot placement and resource allocation strongly affect sensing performance \cite{zhao2023reference}.
Under moderate Doppler, pilot phase progressions over frequency and slow time support delay and velocity estimation. At higher mobility, however, Doppler-induced loss of subcarrier orthogonality produces severe inter-carrier interference (ICI) for OFDM ISAC systems\cite{keskin2021ici}. Although OFDM remains a practical reference, alternative waveforms are worth studying for delay--Doppler sensing in doubly dispersive channels, especially at high mobility \cite{rou2024comparative}.

Affine frequency division multiplexing (AFDM) is one such candidate. AFDM employs the discrete affine Fourier transform (DAFT) and a chirp basis whose parameters can be adapted to the channel delay and Doppler \cite{bemani2023afdm}. This construction yields a controllable affine-domain channel structure and can separate propagation paths by delay and Doppler, providing robustness at high mobility \cite{rou2024comparative,yin2025adaptive}. These properties also motivate pilot-based AFDM sensing. A single DAFT-domain pilot with guards has been used for channel estimation \cite{yin2022pilot} and target sensing \cite{bemani2024isac}. Further pilot designs have addressed sensing ambiguity and channel estimation \cite{zhang2026afdm,ni2025isac}. Other studies have analyzed the ambiguity functions of OFDM and AFDM under normalized signaling models \cite{rou2025ambiguity}.

Nevertheless, existing studies do not establish a fair pilot-based sensing comparison between OFDM and AFDM, as the two waveforms are typically evaluated under different pilot placements and processing assumptions. A systematic comparison under the same pilot resources and otherwise consistent sensing conditions is still lacking, making it difficult to assess their relative performance and guide waveform selection across different operating scenarios. Resolving this issue requires separating the effects of the waveform itself from those of pilot placement and receiver design.

To this end, a resource-matched performance comparison is conducted in this paper for pilot-based sensing with OFDM and AFDM. This paper makes three main contributions. First, it establishes comparable OFDM and AFDM pilot configurations by matching bandwidth, frame duration, total pilot energy, and reserved-resource overhead. Second, it compares random, fixed-comb, and shifted-comb OFDM pilots with AFDM pilots surrounded by guard resources through both aperiodic ambiguity functions without cyclic prefix/chirp-periodic prefix (CP/CPP) and cyclic ambiguity functions with CP/CPP. Third, it evaluates multi-target detection performance in a common unambiguous region, and then studies high-mobility sensing by comparing conventional 2D-FFT and time-domain matched-filter receivers with an AFDM-DAFT receiver adapted to pilot-based sensing, including computational complexity. %Extensive simulations are further carried out to compare their detection performance under different conditions. 

\section{System Model and Pilot Placement}
\subsection{OFDM and AFDM Signal Models}
Consider a single-antenna monostatic ISAC system. The useful time-domain OFDM symbol is
\begin{equation}
\mathbf{x}^{\rm O}_m=\mathbf{F}^{H}\mathbf{s}^{\rm O}_m,
\label{eq:ofdm_tx}
\end{equation}
where $\mathbf{x}^{\rm O}_m$ and $\mathbf{s}^{\rm O}_m$ are the length-$N$ time- and frequency-domain vectors, $\mathbf{F}$ is the unitary discrete Fourier transform (DFT) matrix, and $(\cdot)^H$ denotes the Hermitian transpose. Each frame contains $M$ symbols indexed by $m=0,\ldots,M-1$, each with $N$ subcarriers and useful samples. The sample interval and prefix-inclusive symbol duration are $T_{\rm s}=1/(N\Delta f)$ and $T_{\rm sym}=(N+N_{\rm g})T_{\rm s}$, where $\Delta f$ is the subcarrier spacing and $N_{\rm g}$ is the prefix length in samples. Delays are assumed to be within the CP duration, excluding inter-symbol interference (ISI). 
%The CP is inserted to ensure delays wrap around the prefix after CP removal.

AFDM generates the time-domain symbol using the inverse DAFT,
\begin{equation}
\mathbf{x}^{\rm A}_m=\mathbf{A}^{H}\mathbf{s}^{\rm A}_m,\qquad
\mathbf{A}=\boldsymbol{\Lambda}_{c_2}\mathbf{F}
\boldsymbol{\Lambda}_{c_1},
\label{eq:afdm_tx}
\end{equation}
where $\mathbf{x}^{\rm A}_m$ and $\mathbf{s}^{\rm A}_m$ are the time- and DAFT-domain vectors, and $\mathbf{A}$ is the DAFT matrix. Its diagonal chirp matrices are $\boldsymbol{\Lambda}_{c_i}=\operatorname{diag}\{e^{-j2\pi c_i n^2}\}_{n=0}^{N-1}$, where $n$ is the sample index, $i\in\{1,2\}$, and $c_1,c_2$ are chirp parameters \cite{bemani2023afdm}. The CPP preserves the DAFT-domain channel structure if the delay is within the CPP duration.

Both waveforms share the target channel. For delays within the prefix, the post-prefix received vector is
\begin{equation}
\mathbf{y}_m=\sum_{p=1}^{P}h_p e^{j2\pi f_{{\rm D},p}mT_{\rm sym}}
\mathbf{D}(\nu_p)\boldsymbol{\Pi}^{\ell_p}\mathbf{x}_m
+\mathbf{w}_m,
\label{eq:common_channel}
\end{equation}
where $\mathbf{y}_m$ and $\mathbf{x}_m$ are the received and transmitted vectors for either waveform, and $\mathbf{w}_m$ is additive white Gaussian noise (AWGN). For the $p$th of $P$ targets, $h_p$, $\ell_p$, and $f_{{\rm D},p}$ denote the complex gain, delay in samples, and Doppler frequency, with normalized Doppler $\nu_p=f_{{\rm D},p}/\Delta f$. The target subscript is omitted in the single-target notation $\nu$. The operator $\boldsymbol{\Pi}^{\ell_p}$ applies a circular delay of $\ell_p$ samples.
%; for integer delays, this also represents CPP processing with the parameters in Table~\ref{tab:simulation_parameters}. 
The within-symbol Doppler operator is
\begin{equation}
\mathbf{D}(\nu_p)=\operatorname{diag}
\{e^{j2\pi\nu_p n/N}\}_{n=0}^{N-1}.
\end{equation}
Thus, \eqref{eq:common_channel} retains within-symbol Doppler variation and the resulting ICI.

Applying the receive DFT to OFDM gives
\begin{equation}
\mathbf{Y}^{\rm O}_m=\sum_{p=1}^{P}h_p e^{j2\pi f_{{\rm D},p}mT_{\rm sym}}
\underbrace{\mathbf{F}\mathbf{D}(\nu_p)
\boldsymbol{\Pi}^{\ell_p}\mathbf{F}^{H}}_{\mathbf{H}^{\rm O}_p}
\mathbf{s}^{\rm O}_m+\mathbf{W}^{\rm O}_m,
\label{eq:ofdm_io}
\end{equation}
where $\mathbf{Y}^{\rm O}_m=\mathbf{F}\mathbf{y}_m$ and $\mathbf{W}^{\rm O}_m=\mathbf{F}\mathbf{w}_m$ are the frequency-domain received vector and noise, and $\mathbf{H}^{\rm O}_p$ is the effective channel of target $p$. For small $\nu_p$, $\mathbf{H}^{\rm O}_p$ is approximately diagonal, giving separable phase progressions over subcarriers and symbols. At high Doppler, off-diagonal terms cause ICI.

Applying the DAFT to AFDM instead yields
\begin{equation}
\mathbf{Y}^{\rm A}_m=\sum_{p=1}^{P}h_p e^{j2\pi f_{{\rm D},p}mT_{\rm sym}}
\underbrace{\mathbf{A}\mathbf{D}(\nu_p)
\boldsymbol{\Pi}^{\ell_p}\mathbf{A}^{H}}_{\mathbf{H}^{\rm A}_p}
\mathbf{s}^{\rm A}_m+\mathbf{W}^{\rm A}_m,
\label{eq:afdm_io}
\end{equation}
where $\mathbf{Y}^{\rm A}_m=\mathbf{A}\mathbf{y}_m$ and $\mathbf{W}^{\rm A}_m=\mathbf{A}\mathbf{w}_m$ are the DAFT-domain received vector and noise, and $\mathbf{H}^{\rm A}_p$ is the effective channel of target $p$. With suitable chirp parameters, delay and integer Doppler determine a localized DAFT-domain shift, while fractional Doppler spreads energy around it \cite{bemani2023afdm,ni2025isac}.

\begin{figure}[t]
\centering
\includegraphics[width=\columnwidth]{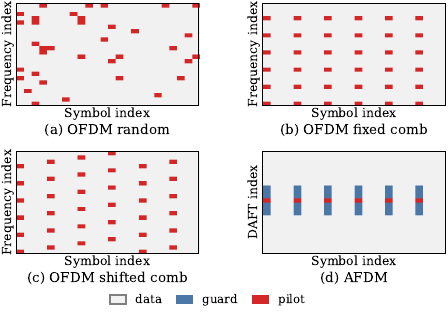}
\caption{Pilot placement schemes. OFDM pilots occupy the frequency--time grid, whereas AFDM uses single pilots with guard resources in the DAFT--time grid.}
\label{fig:pilot_schemes}
\vspace{-2mm}
\end{figure}

\subsection{Pilot Placement Schemes}
For either waveform, write the modulation-domain frame as
\begin{equation}
\mathbf{S}=\mathbf{P}+\mathbf{D}_{\rm data},\qquad
\operatorname{supp}(\mathbf{P})\cap
\operatorname{supp}(\mathbf{D}_{\rm data})=\varnothing,
\end{equation}
where $\mathbf{S}$ is the $N\times M$ modulation-domain frame, and $\mathbf{P}$ and $\mathbf{D}_{\rm data}$ contain known pilots and unknown data, respectively. Here $P_{k,m}$ denotes the $(k,m)$ entry of $\mathbf{P}$, and $\operatorname{supp}(\cdot)$ denotes support. Figure~\ref{fig:pilot_schemes} illustrates the four pilot placement schemes.

For \emph{OFDM-random}, a fixed number of pilot positions is selected randomly over the frequency--time grid. For \emph{OFDM-fixed-comb}, the pilot positions form the set
\begin{equation}
\Omega_{\rm comb}=\{(k,m): k=k_0+rD_{\rm f},\ m=m_0+sD_{\rm t}\},
\end{equation}
where $k$ indexes subcarriers, $(k_0,m_0)$ is the starting position, and $D_{\rm f},D_{\rm t}$ are the spacings in subcarriers and symbols. Nonnegative integers $r,s$ enumerate positions within $0\leq k<N$ and $0\leq m<M$. For \emph{OFDM-shifted-comb}, the subcarrier positions are
\begin{equation}
k=\langle k_0+\delta_m+rD_{\rm f}\rangle_N,
\end{equation}
where $\delta_m$ is the integer comb shift and $\langle\cdot\rangle_N$ denotes modulo-$N$ indexing. The shift changes the range--Doppler ambiguity pattern while preserving the pilot-bearing symbol indices.

%The \emph{AFDM} scheme uses one pilot per pilot-bearing symbol, with surrounding zero-valued DAFT-domain guards separating pilot and data responses over the assumed delay-Doppler region. 

For \emph{AFDM} scheme, we adopt the established single-pilot-with-guards configuration\cite{bemani2024isac}. Through the inverse DAFT, a single pilot generates a chirp waveform whose delay-Doppler response can be exploited for sensing, without requiring multiple nonzero pilots per symbol. Surrounding zero-valued guards are reserved to mitigate interference from unknown data.
%The scheme provides a representative single-pilot sensing baseline under the same resource condition.

\section{Ambiguity Function Analysis and Results}
\subsection{Ambiguity Function}
For pilot-based sensing, only known pilots are utilized for sensing processing. For simplicity, we assume the data are zero. Let $p_m[n]$ be the pilot waveform at sample $n=0,\ldots,N-1$ of symbol $m=0,\ldots,M-1$, and $E_{\rm p}=\sum_{m,n}|p_m[n]|^2$ the total useful-sample pilot energy per frame. 

\subsubsection{Aperiodic Ambiguity Functions without CP/CPP}
The concatenated useful symbols form $p[n]$, zero outside $0\leq n<MN$, with normalized aperiodic ambiguity function
\begin{equation}
\chi_{\rm ap}(\ell,f)=\frac{1}{E_{\rm p}}\sum_n p[n]p^*[n-\ell]
e^{-j2\pi fn/B},\quad B=N\Delta f,
\label{eq:aperiodic_af}
\end{equation}
where $\ell$ is integer sample delay, $f$ Doppler frequency, $B$ bandwidth, and $(\cdot)^*$ denotes complex conjugation. Range and velocity are $R=c\ell/(2B)$ and $v=cf/(2f_{\rm c})$, where $c$ is propagation speed and $f_{\rm c}$ is carrier frequency.

For OFDM-fixed-comb with $D_{\rm f}$ dividing $N$, the delay response repeats every $N/D_{\rm f}$ samples, giving the maximum unambiguous range
\begin{equation}
R_{\rm u}=\frac{c}{2D_{\rm f}\Delta f}.
\label{eq:range_alias_structure}
\end{equation}
The full-frequency-grid reference is $R_{\rm u}=c/(2\Delta f)$. These limits do not determine aperiodic peak heights. Similarly, pilots every $D_{\rm t}$ symbols give the conventional slow-time maximum unambiguous velocity,
\begin{equation}
v_{\rm u}=\frac{c}{4f_{\rm c}D_{\rm t}T},\qquad T=NT_{\rm s},
\label{eq:pilot_velocity_interval}
\end{equation}
where $T$ is the useful-symbol duration, with principal interval $|v|<v_{\rm u}$. Irregular placement generally produces a distributed ambiguity background instead of regular peaks.

\subsubsection{Cyclic Ambiguity Functions with CP/CPP}
After prefix removal, the CP and CPP delay operators are
\begin{align}
\mathcal{T}^{\rm CP}_{\ell}(p_m[n])
&=p_m[\langle n-\ell\rangle_N],\\
\mathcal{T}^{\rm CPP}_{\ell}(p_m[n])
&=p_m[\langle n-\ell\rangle_N]
e^{-j2\pi c_1(\langle n-\ell\rangle_N^2-(n-\ell)^2)}.
\end{align}
With the corresponding $\mathcal{T}_{\ell}$, define
\begin{equation}
\begin{aligned}
\chi_{\rm cyc}(\ell,f)
&=\frac{1}{E_{\rm p}}\sum_{m=0}^{M-1}\sum_{n=0}^{N-1}
\mathcal{T}_{\ell}(p_m[n])p_m^*[n]\\
&\quad\times e^{-j2\pi f[n+m(N+N_{\rm g})]/B}.
\end{aligned}
\label{eq:cyclic_af}
\end{equation}
With the prefix-inclusive interval $T_{\rm sym}=(N+N_{\rm g})/B$, the maximum unambiguous velocity becomes $v_{\rm u}=c/(4f_{\rm c}D_{\rm t}T_{\rm sym})$.

For one AFDM pilot per pilot-bearing symbol at fixed DAFT index $k_{\rm p}$, the inverse DAFT gives $p_m[n]=(a_m/\sqrt{N})e^{j2\pi(c_1n^2+k_{\rm p}n/N)}$, where $a_m$ is the complex pilot coefficient including the constant chirp phase, and is zero in symbols without pilots. Substituting the CPP shift yields
\begin{equation}
|\chi_{\rm cyc}(\ell,0)|=
\left|\frac{1}{N}\sum_{n=0}^{N-1}e^{-j2\pi q\ell n/N}\right|,
\qquad q=2Nc_1.
\label{eq:afdm_cyclic_slice}
\end{equation}
For integer $q$ coprime to $N$, the sum gives
\begin{equation}
|\chi_{\rm cyc}(\ell,0)|=0,\qquad \ell=1,\ldots,N-1.
\label{eq:afdm_zero_range}
\end{equation}
These zeros are independent of pilot amplitudes and pilot-bearing symbol count, but require integer delays and correct CPP processing. They do not imply zero sidelobes at fractional delays or nonzero Doppler.

\subsection{Typical Case Study}
\begin{table}[!t]
\caption{Common simulation parameters}
\label{tab:simulation_parameters}
\centering
\footnotesize
\setlength{\tabcolsep}{2pt}
\renewcommand{\arraystretch}{0.94}
\begin{tabular}{@{}lc@{\hspace{16pt}}lc@{}}
\toprule
Parameter & Value & Parameter & Value\\
\midrule
Carrier $f_{\rm c}$ & 77 GHz & Spacing $\Delta f$ & 120 kHz\\
Subcarriers $N$ & 256 & Symbols $M$ & 64\\
Bandwidth $B$ & 30.72 MHz & Sample time $T_{\rm s}$ & 32.55 ns\\
CP/CPP length $N_{\rm g}$ & 32 & $T_{\rm sym}$ & 9.375 $\mu$s\\
$c_1$ & $3/512$ & $c_2$ & $\pi$\\
Time spacing $D_{\rm t}$ & $4$ & Frequency spacing $D_{\rm f}$ & $8$\\
\bottomrule
\end{tabular}
\vspace{-2mm}
\end{table}

Using \eqref{eq:aperiodic_af} and \eqref{eq:cyclic_af}, a typical case compares AFDM and OFDM through ambiguity function, peak-to-sidelobe ratio (PSLR), and integrated sidelobe-to-mainlobe energy ratio (ISLR). These metrics depend on frame size, pilot design, chirp parameters, evaluation grid, and mainlobe convention.

Using Table~\ref{tab:simulation_parameters}, each scheme reserves 512 resources. For OFDM-random, pilots are randomly placed within 64 symbols. For OFDM with structured pilot placement, 16 pilot symbols are used, and each pilot symbol contains 32 QPSK-modulated pilots. AFDM uses the same pilot symbols, each with one DAFT-domain pilot and 31 neighboring zero guards. The total energy of the pilot for all the schemes is set to be the same.

%OFDM-random has the same total $E_{\rm p}=16000E_{\rm d}$, where $E_{\rm d}$ is the energy per data resource. Structured schemes assign $1000E_{\rm d}$ per pilot-bearing symbol. 

%Figure~\ref{fig:ambiguity_comparison} displays $10\log_{10}(|\chi(\ell,f)|^2/|\chi(0,0)|^2)$, clipped to $[-60,0]$~dB, where $\chi$ denotes either ambiguity function. Figure~\ref{fig:pilot_only_range_slice} shows the aperiodic zero-Doppler range cuts: PSLR compares the peak with the highest sidelobe, whereas ISLR measures total sidelobe energy relative to mainlobe energy. Table~\ref{tab:ambiguity_metrics} reports aperiodic/cyclic pilot-only metrics in centered windows with unambiguous range and velocity. Range and Doppler PSLRs use the axial cuts and 2D metrics use the full map within the corresponding window. Higher PSLR and lower ISLR are preferable.

\begin{figure*}[t]
\centering
\subfloat[Without CP/CPP: aperiodic ambiguity functions.\label{fig:ambiguity_aperiodic}]{%
\includegraphics[width=0.485\textwidth]{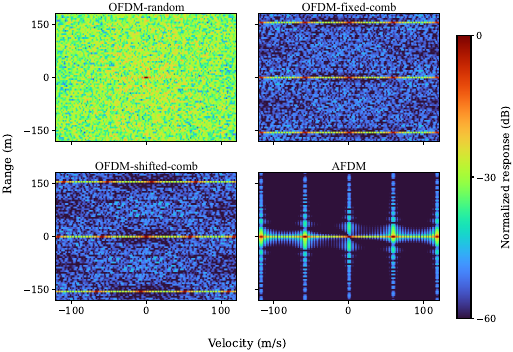}}
\hfill
\subfloat[With CP/CPP: cyclic ambiguity functions.\label{fig:ambiguity_cyclic}]{%
\includegraphics[width=0.485\textwidth]{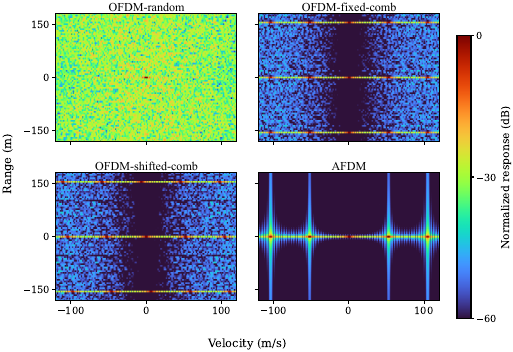}}
\caption{Pilot-only ambiguity comparison under the same 512 reserved resources, total pilot energy, and constant-modulus QPSK pilot rule. In (b), only the CP/CPP-supported range is physically free of inter-symbol interference.}
\label{fig:ambiguity_comparison}
\vspace{-2mm}
\end{figure*}

\begin{figure}[!t]
\centering
\includegraphics[width=\columnwidth,height=0.42\columnwidth,keepaspectratio]{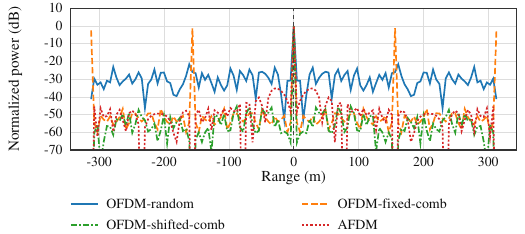}
\caption{Pilot-only aperiodic ambiguity: zero-Doppler range slices for the four pilot schemes.}
\label{fig:pilot_only_range_slice}
\vspace{-2mm}
\end{figure}

\begin{comment}
\begin{table}[t]
\caption{Pilot-only ambiguity metrics, reported as aperiodic/cyclic}
\label{tab:ambiguity_metrics}
\centering
\footnotesize
\setlength{\tabcolsep}{2pt}
\renewcommand{\arraystretch}{1.02}
%\begin{tabular}{\textwidth}{@{\extracolsep{\fill}}lcccc@{}}
\begin{tabular}{@{}lcccc@{}}
\toprule
Pilot scheme & Range PSLR (dB) & Doppler PSLR (dB) & 2D PSLR (dB) & 2D ISLR (dB) \\
\midrule
OFDM-random & 21.58/20.48 & 13.83/13.83 & 13.83/13.83 & 11.39/13.00 \\
OFDM-fixed-comb & 1.16/0.00 & 0.91/0.72 & 0.91/0.00 & 12.02/13.33 \\
OFDM-shifted-comb & 45.40/$\geq150^\dagger$ & 0.91/0.72 & 0.91/0.01 & 12.03/13.33 \\
AFDM & 35.24/$\geq150^\dagger$ & 0.91/0.72 & 0.91/0.72 & 5.02/5.66 \\
\bottomrule

\multicolumn{5}{@{}l}{\scriptsize $^\dagger$ No sidelobe exceeds the $-150$ dB numerical floor in the evaluated zero-Doppler slice.}
\end{tabular}
\vspace{-1mm}
\end{table}
\end{comment}

Figure~\ref{fig:ambiguity_comparison} shows the ambiguity results of the four pilot schemes, which are displayed by normalized response $10\log_{10}(|\chi(\ell,f)|^2/|\chi(0,0)|^2)$, clipped to $[-60,0]$~dB. For OFDM-fixed-comb, uniformly spaced subcarriers produce range ambiguity peaks, corresponding to an unambiguous range of $156.25$~m. OFDM-shifted-comb introduces delay phases that suppress these peaks on the zero-Doppler cut but move them to coupled delay-Doppler positions. AFDM achieves range compression with a single DAFT-domain pilot through chirp matching, without the comb-induced range ambiguity peaks. With CP, the OFDM ambiguity functions exhibit an elevated sidelobe background as velocity increases, because intra-symbol Doppler disrupts subcarrier orthogonality. With CPP, \eqref{eq:afdm_zero_range} gives AFDM zero sidelobes at integer delays along the zero-Doppler cut. OFDM-random disrupts regular phase alignment, weakening periodic peaks but leaving a distributed ambiguity background. Despite their different range responses, OFDM-fixed-comb, OFDM-shifted-comb, and AFDM share the same symbol spacing $D_{\rm t}=4$, resulting in periodic Doppler ambiguities caused by sparse slow-time sampling.

Figure~\ref{fig:pilot_only_range_slice} details the zero-Doppler range response in Fig.~\ref{fig:ambiguity_aperiodic}. OFDM-fixed-comb exhibits low sidelobes between strong periodic range peaks. OFDM-shifted-comb suppresses these repeated peaks on the range cut, while AFDM forms a distinct main peak with residual chirp sidelobes. OFDM-random instead produces a higher distributed sidelobe background.

Table~\ref{tab:ambiguity_metrics} compares sensing performance through PSLR and ISLR. The PSLR compares the peak with the highest sidelobe, whereas the ISLR measures total sidelobe energy relative to mainlobe energy. All four schemes use centered comparison windows with $|R|<78.125$~m and $|v|<29.22$~m/s without CP/CPP or $|v|<25.97$~m/s with CP/CPP. Without CP/CPP, OFDM-fixed-comb and OFDM-shifted-comb achieve higher range PSLRs than AFDM within the windows. For OFDM-fixed-comb, the aperiodic correlation retains residual sidelobes caused by truncation at the symbol boundaries. Their magnitudes depend on the QPSK pilot phases and may be reduced by phase cancellation across symbols. In contrast, the single-pilot AFDM waveform produces a chirp with deterministic aperiodic correlation sidelobes, since the common QPSK pilot phase cancels in the autocorrelation. Its largest range sidelobe occurs at a delay of six samples, yielding a PSLR of $35.24$~dB. With CP/CPP, cyclic correlation removes the boundary-truncation effect, and all three structured schemes reach the numerical floor for sampled range sidelobes within the windows. Nevertheless, their 2D PSLRs remain near $13.30$~dB, matching their Doppler PSLRs.

\begin{table}[t]
\caption{Pilot-only ambiguity metrics in the centered comparison windows (aperiodic/cyclic)}
\label{tab:ambiguity_metrics}
\centering
\footnotesize
\setlength{\tabcolsep}{2pt}
\renewcommand{\arraystretch}{1.02}
\resizebox{\columnwidth}{!}{%
\begin{tabular}{@{}lcccc@{}}
\toprule
Pilot scheme
& \makecell{Range PSLR\\(dB)}
& \makecell{Doppler PSLR\\(dB)}
& \makecell{2D PSLR\\(dB)}
& \makecell{2D ISLR\\(dB)} \\
\midrule
OFDM-random & 23.50/23.52 & 13.83/13.83 & 13.83/13.83 & 0.15/0.39 \\
OFDM-fixed-comb & 46.76/$\geq150^{\dagger}$ & 13.30/13.30 & 13.30/13.30 & -9.67/-9.77 \\
OFDM-shifted-comb & 45.80/$\geq150^{\dagger}$ & 13.30/13.29 & 13.30/13.29 & -9.66/-9.76 \\
AFDM & 35.24/$\geq150^{\dagger}$ & 13.30/13.30 & 13.30/13.30 & -9.64/-9.77 \\
\bottomrule
\multicolumn{5}{@{}l}{\scriptsize $^{\dagger}$ Range sidelobes are below the $-150$ dB numerical floor.} \\
\end{tabular}
}
\vspace{-1mm}
\end{table}

OFDM-fixed-comb, OFDM-shifted-comb, and AFDM have similar 2D ISLRs in both analyses. OFDM-random instead distributes residual correlations across many delay-Doppler cells. Their accumulated energy gives approximately $10$~dB higher ISLR, even without strong regular repetitions. Such distributed interference may mask weak target echoes in the presence of stronger targets. These results show that structured pilot placement concentrates more energy within the mainlobe, reducing the overall sidelobe background. 

\begin{figure*}[t]
\centering
\includegraphics[width=0.94\textwidth]{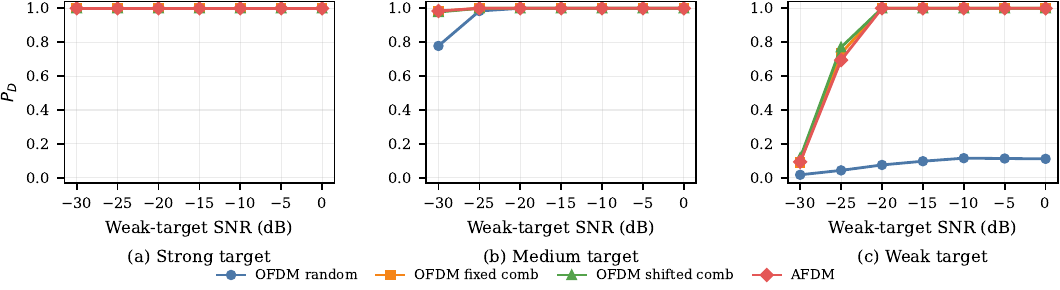}
\caption{Per-target detection probability in the three-target scene per weak-target SNR: (a) strong target, (b) medium target, and (c) weak target. }
\label{fig:multi_target}
\end{figure*}

\begin{remark}[Pilot placement and two-dimensional ambiguity]
Under matched resources, pilot placement changes coherent repetition, delay--Doppler coupling, and sidelobe energy distribution. OFDM-fixed-comb produces range ambiguity peaks, OFDM-shifted-comb relocates them across delay and Doppler and AFDM exhibits a distinct chirp-based response. Structured OFDM and AFDM achieve similar 2D PSLR and ISLR, whereas OFDM-random shows a higher ISLR. Thus, favorable peak sidelobe performance does not necessarily imply low integrated interference or ambiguity suppression, motivating joint evaluation of ambiguity, PSLR and ISLR.
\end{remark}

\section{Target Detection Results}
\subsection{Signal Processing and Complexity}
All receivers use known pilots as matching references. OFDM applies a DFT to received symbols and multiplies them by the conjugate pilots. A subcarrier inverse FFT (IFFT) and slow-time FFT produce the range-Doppler power map. The AFDM time-domain matched-filter receiver uses FFT/IFFT-based circular correlation, followed by a slow-time FFT.

For high mobility, the AFDM-DAFT receiver adapts the AFT--Doppler method in \cite{ni2025isac} to pilot sensing. It matches DAFT-domain pilot responses over candidate delays and affine-shift branches, then applies slow-time FFTs. Combining delay-dependent DAFT displacement with the fractional-Doppler estimate resolves integer Doppler beyond the conventional maximum unambiguous velocity.

With reference transforms precomputed, conventional receivers have comparable FFT-based map-generation costs,
\begin{equation}
C_{\rm conv}\approx MN(2\log_2N+\log_2M+1).
\end{equation}
The pilot-based AFDM-DAFT receiver uses symbolwise DAFTs and $AL$ slow-time FFTs, giving
\begin{equation}
C_{\rm pilot}=\mathcal{O}\!\left(MN\log_2N+ALM\log_2M\right),
\end{equation}
where $L$ counts delay hypotheses and $A=2\alpha_{\max}+1$ counts affine branches, with maximum absolute index $\alpha_{\max}$. Its $ALM$-cell detection grid, versus $LM$ for conventional processing, increases search costs for detection.

Power maps use two-dimensional cell-averaging constant-false-alarm-rate (CA-CFAR) detection and local-maximum extraction. Detection probability $P_{\rm D}$ is the fraction of trials with a candidate inside the association gate and multi-target matching is one-to-one. Estimates use integer bins without interpolation.

%Experiments use Table~\ref{tab:simulation_parameters}, equal frame pilot energy and reserved-resource overhead across schemes, and unknown QPSK data on non-reserved resources. 
For target sensing in low-mobility scenarios, the pilot placement and simulation parameters are the same as the case study in the previous section. 
For high-mobility sensing, 2048 resources are used, with pilots allocated in every symbol. Each point uses 500 trials, per-cell false-alarm probability $P_{\rm FA}=10^{-4}$, a $9\times9$ CA-CFAR window with a $3\times3$ guard region, and a $\pm1.5$-bin range-velocity association gate. Echo sample signal-to-noise ratio (SNR), the ratio of mean target-echo sample power to noise power, spans $-30$ to $0$~dB unless stated otherwise. 

\begin{comment}

\subsection{Single-Target Sensing}
\begin{figure*}[t]
\centering
\includegraphics[width=0.94\textwidth]{fig3_single_target_results.pdf}
\caption{Single-target sensing per SNR: (a) target detection probability, (b) range RMSE over all trials, and (c) velocity RMSE over all trials.}
\label{fig:single_target}
\end{figure*}
To avoid ambiguity peak interference, each trial draws an independent target from $R\sim\mathcal{U}(0,150)$ m and $v\sim\mathcal{U}(-20,20)$ m/s, where $\mathcal{U}$ denotes a uniform distribution, with independent continuous range and velocity in the common unambiguous region. Figure~\ref{fig:single_target} reports $P_{\rm D}$ and range and velocity root-mean-square error (RMSE) over all trials. %Estimates select the strongest CFAR candidate, or the strongest peak in the common search region if none passes.

Detection curves in Fig.~\ref{fig:single_target}(a) are similar. At $-25$~dB, $P_{\rm D}$ is $0.798$, $0.806$, $0.824$, and $0.806$ for OFDM-random, OFDM-fixed-comb, OFDM-shifted-comb, and AFDM, respectively. At $-20$~dB, all are at least $0.998$. Below the detection transition, misses and location errors dominate RMSEs in Fig.~\ref{fig:single_target}(b)-(c). At $-25$~dB, range and velocity RMSEs are $11.61$--$17.01$~m and $3.68$--$5.30$~m/s. At $-20$~dB, they fall to about $1.45$~m and $1.06$~m/s. Higher SNR gives little improvement because estimates use integer range-Doppler bins. With equal pilot energy in the common unambiguous region, all schemes have similar detection thresholds and no clear accuracy advantage.

\end{comment}

\subsection{Target Sensing in Low-Mobility Scenarios}

In this experiment, target detection is evaluated under a multi-target setup, with a representative three-target scenario presented here. The strong, medium, and weak targets are at $(40.0,-13.9)$, $(120.1,15.7)$, and $(69.3,-4.7)$ on the range-velocity map, respectively, with relative powers of $16$, $8$, and $0$~dB. The weak-target SNR varies from $-30$ to $0$~dB, while the other target SNRs maintain their respective offsets.

Figure~\ref{fig:multi_target} reports per-target detection. All schemes can detect the strong target every trial. At a weak-target SNR of $-30$~dB, the medium-target detection probabilities are $0.778$ for OFDM-random and approximately $0.98$ for the three structured schemes. At $-25$~dB, they rise to $0.984$ for OFDM-random and 1 for the other schemes. The weak target shows the clearest difference: all structured placements reach one by $-20$~dB, while OFDM-random remains between $0.076$ and $0.116$ from $-20$ to $0$~dB. Structured placement improves weak-target detection, whereas the distributed background of random placement masks the weak echo in this scene.

\begin{remark}[Interference-limited weak-target detection]
Structured placement for both OFDM and AFDM gives reliable weak-target detection as SNR increases, whereas random placement for OFDM remains poor despite lower peak ambiguity sidelobes. With fixed target power ratios, reducing noise leaves relative strong-target interference unchanged. Detection can remain interference-limited at high SNR, making sidelobes at target locations a key pilot-design criterion.
\end{remark}

\subsection{Target Sensing in High-Mobility Scenarios}
\begin{figure}[t]
\centering
\includegraphics[width=0.98\columnwidth]{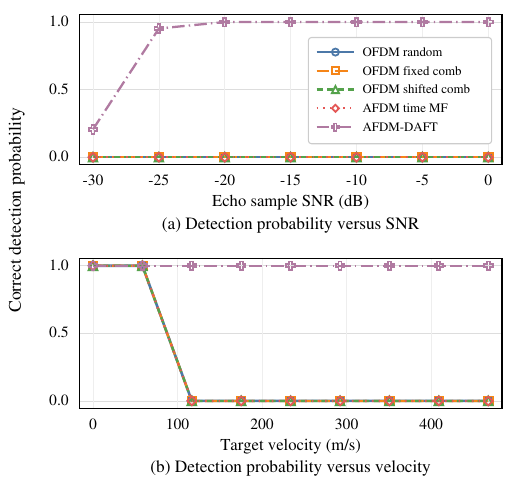}
\caption{High-mobility detection at $R=50$~m with 500 trials per point: (a) detection probability versus SNR at $\nu=1.5$, (b) detection probability versus target velocity at $0$~dB.}
\label{fig:high_mobility}
\vspace{-2mm}
\end{figure}

The dense-pilot experiment compares three OFDM receivers, the AFDM time-domain matched-filter receiver, and AFDM-DAFT for a target at $R=50$~m. Figure~\ref{fig:high_mobility}(a) varies SNR at normalized Doppler $\nu=1.5$ ($v=350.649$~m/s), while Fig.~\ref{fig:high_mobility}(b) varies $\nu$ from 0 to 2 ($v=0$ $\sim$ $467.532$~m/s) at $0$~dB.

With pilots in every symbol, the conventional maximum unambiguous velocity is $v_{\rm u}=103.896~\mathrm{m/s}$.
At $\nu=1.5$, all conventional receivers return aliased velocities outside the target gate, giving zero correct detection probability across SNRs. In contrast, AFDM-DAFT exploits the delay-dependent DAFT-domain displacement to identify integer Doppler branches, enabling recovery of the true velocity beyond the slow-time limit. AFDM-DAFT instead gives $P_{\rm D}=0.206$ and $0.952$ at $-30$ and $-25$~dB, respectively, and reaches one from $-20$~dB onward. In the velocity sweep, conventional receivers give $P_{\rm D}=1$ until $58.442$~m/s but fail at the first tested velocity beyond the conventional maximum unambiguous velocity. AFDM-DAFT maintains $P_{\rm D}=1$ through $467.532$~m/s.

\begin{remark}[Joint waveform and receiver design]
For high-mobility scenarios, conventional slow-time processing suffers from Doppler ambiguity beyond its unambiguous velocity range for both OFDM and AFDM. In contrast, the AFDM-DAFT receiver exploits the DAFT-domain structure to resolve integer Doppler shifts beyond this limit, at the cost of increased search complexity.
\end{remark}

\section{Conclusion}
This paper compared the pilot-based sensing performance of AFDM and OFDM through ambiguity-function analyses and target sensing experiments under different mobilities. OFDM-fixed-comb exhibits low range sidelobes but periodic range ambiguity peaks, while OFDM-shifted-comb relocates these peaks to coupled delay-Doppler positions. AFDM achieves range compression with a single DAFT-domain pilot per symbol. Within the evaluated unambiguous windows, OFDM-fixed-comb, OFDM-shifted-comb and AFDM have similar 2D ISLRs, about $10$~dB below OFDM-random. In the common unambiguous region, all schemes have similar detection probability for strong-target detection. However, without interference cancellation, the three structured sparse pilot placements reliably detect the weak target from $-20$~dB onward, whereas OFDM-random is limited by its distributed correlation background. For high mobility, OFDM 2D-FFT and AFDM time-domain matched-filter receivers lose the physical Doppler branch beyond the maximum unambiguous velocity, whereas AFDM-DAFT exploits the waveform's structured DAFT-domain displacement and maintains a high detection probability at high speed. This extension requires processing that exploits the DAFT structure and comes with a larger detection grid.

%With sparse pilots, AFDM and OFDM-shifted-comb suppress range sidelobes relative to OFDM-fixed-comb, but retain Doppler ambiguity peaks from symbol-level slow-time sampling. 

\bibliographystyle{IEEEtran}
\bibliography{references}

\end{document}